\documentclass[preprint,aps,nofootinbib]{revtex4}

\usepackage{graphicx}

\usepackage{amsmath}
\usepackage{amsfonts}
\usepackage{amssymb}
\usepackage{color}

\def\be{\begin{equation}}
\def\ee{\end{equation}}
\def\bea{\begin{eqnarray}}
\def\eea{\end{eqnarray}}

\def\slashchar#1{\setbox0=\hbox{$#1$}           
   \dimen0=\wd0                                 
   \setbox1=\hbox{/} \dimen1=\wd1               
   \ifdim\dimen0>\dimen1                        
      \rlap{\hbox to \dimen0{\hfil/\hfil}}      
      #1                                        
   \else                                        
      \rlap{\hbox to \dimen1{\hfil$#1$\hfil}}   
      /                                         
   \fi}

\begin{document}

\vspace*{1cm}

\title{A Minimally Symmetric Higgs Boson}

\author{\vspace{0.5cm} Ian Low}
\affiliation{\vspace{0.5cm}
\mbox{High Energy Physics Division, Argonne National Laboratory, Argonne, IL 60439}\\
\mbox{Department of Physics and Astronomy, Northwestern University, Evanston, IL 60208} \\
 \vspace{0.5cm}
}

\begin{abstract}
\vspace{1cm}
Models addressing the naturalness  of a light Higgs boson typically employ  symmetries, either bosonic or fermionic, to stabilize the Higgs mass. We consider a setup with the minimal amount of symmetries: four  shift symmetries acting on the four components of the Higgs doublet, subject to the constraints of linearly realized $SU(2)_L\times U(1)_Y$ electroweak symmetry.  Up to terms that explicitly violate the shift symmetries, the effective lagrangian can be derived, irrespective of the spontaneously broken group $G$ in the ultraviolet, and is universal among all models where the Higgs arises as a pseudo-Nambu-Goldstone boson (PNGB). Very high energy scatterings of vector bosons could provide smoking gun signals of a minimally symmetric Higgs boson. 

\end{abstract}

\maketitle

\section{Introduction}
\label{sect:intro}

The discovery of a light Higgs boson at around 125 GeV sharpens the question of naturalness: what is the mechanism that stabilizes the light Higgs mass? There are two popular classes of solutions: one invokes fermionic symmetry acting on the Higgs, which is supersymmetry, while the other uses bosonic symmetry, in which case the Higgs is considered a PNGB. The latter possibility is the focus of this work.

The history of a PNGB Higgs goes back to Refs.~\cite{Kaplan:1983fs, Kaplan:1983sm}, and it gained new popularity after the proposal of little Higgs theories \cite{ArkaniHamed:2001nc,ArkaniHamed:2002qx,ArkaniHamed:2002qy}. Subsequently the holographic Higgs models \cite{Contino:2003ve}, which is based on the AdS/CFT conjecture \cite{Maldacena:1997re}, and its cousin \cite{Agashe:2004rs} also received much attention. 

Constructions of  models with a PNGB Higgs rely on two seminal papers by  Coleman, Callan, Wess and Zumino (CCWZ) \cite{Coleman:sm,Callan:sn},  which considered all possible nonlinear realizations of a broken group $G$ which become linear when restricting to the unbroken group $H$. The CCWZ construction is top-down, requiring the specification of a particular symmetry breaking pattern $G/H$, before writing the effective lagrangian. The only requirement of $G/H$ is that it contains a PNGB transforming linearly as a doublet under the electroweak $SU(2)_L\times U(1)_Y$ gauge group. As such, there exists an infinite number of possibilities for $G/H$, and CCWZ gives a seemingly different effective lagrangian for each coset $G/H$. Indeed, the review article in Ref.~\cite{Bellazzini:2014yua} lists a large number of possible cosets. The landscape of models with a PNGB Higgs looks huge.

The situation is very different in supersymmetry, where there is a minimal lagrangian contained in all supersymmetric models: the minimally supersymmetric standard model \cite{Dimopoulos:1981zb, Djouadi:2005gj}. There were previous attempts to consider the minimal, universal ingredients of a PNGB Higgs and the resulting lagrangian. Progress was made when the Strongly-Interacting Light Higgs (SILH) lagrangian was proposed in Ref.~\cite{Giudice:2007fh}, which is based on a set of power counting rules to determine which higher dimensional operators are more important in models with a PNGB Higgs. Further progress was achieved when it was discovered that certain operators in the SILH lagrangian carry definitive signs under broad assumptions \cite{Low:2009di}. One example is the coefficient of a certain dimension-six operator involving two derivatives and four Higgses, $c_H$, which must be positive for a compact coset, unless there exists a charge-2 scalars coupling to the Higgs.\footnote{See also the discussion in Ref.~\cite{Bellazzini:2014waa}.}

The purpose of this work is to continue in the direction of understanding universal features among all PNGB Higgs models. Instead of following the top-down approach of CCWZ, we  opt for a bottom-up perspective, by postulating four shift symmetries acting on the four components of the Higgs doublet. By requiring the lagrangian must satisfy  the Adler's zero condition \cite{Adler:1964um}, which states that the scattering amplitudes of Goldstone bosons must vanish when emitting a soft Goldstone, as well as the linearly realized electroweak symmetry, we are able to derive the effective lagrangian without referring to any particular symmetry breaking pattern $G/H$. The Adler's zero condition was initially derived in the context of the pion scattering amplitudes by using the partially conserved axial current (PCAC) hypothesis in low-energy QCD. It forbids a constant term in the pion scattering amplitudes and is often loosely described as the pions are always derivatively coupled. It is well-known that derivative couplings are a universal feature of Goldstone bosons. Our approach will be to consistently combine this condition with the requirement that the Goldstone bosons transform linearly under the electroweak $SU(2)_L\times U(1)_Y$.

The generality of such an approach is studied in a separate publication \cite{Low:shift}. In this work we only focus on the phenomenologically interesting scenario of a PNGB transforming linearly as a doublet under $SU(2)_L\times U(1)_Y$, when there is no $SU(2)_C$ custodial invariance, or as a fundamental representation under the $SO(4)$, which implements the custodial invariance in the Higgs sector  \cite{Sikivie:1980hm} and suppresses the contribution to the electroweak $\rho$ parameter.

One parameter that cannot be determined in the IR is the normalization of the Goldstone decay constant $f$, which is sensitive to the coset structure $G/H$. Otherwise the lagrangian we derive is universal and contained in all PNGB Higgs models, up to terms that explicitly violate the shift symmetries. In particular, Yukawa couplings of standard model (SM) fermions necessarily break the shift symmetry and, as such, their implementation is model-dependent. 

One could consider UV completions of a minimally symmetric Higgs, by adopting a particular symmetry breaking pattern $G/H$. In this context, it is possible to choose a minimal coset that contains the universal lagrangian   derived in this work. One example is the $SO(5)/SO(4)$ model in Ref.~\cite{Agashe:2004rs}, which is the minimal coset giving rise to PNGBs furnishing the fundamental representation of the $SO(4)$ group. Nevertheless, it is important to understand what constraints and predictions of the minimal coset model are universal among all PNGB models and what are not. We will see that some predictions of the minimal coset  are indeed universal, especially when it comes to high energy scatterings of vector bosons.

This work is organized as follows: in the next section we derive the effective lagrangians for a minimally symmetric Higgs boson, considering cases with and without the $SU(2)_C$ custodial invariance. Then we propose a simple bottom-up implementation of Yukawa couplings in the top sector, which captures many features studied in coset constructions. Experimental constraints and collider signals are considered in the next Section, which is followed by the conclusion.

\section{The effective lagrangians}

In this section we derive the effective lagrangians for a minimally symmetric Higgs boson transforming as a fundamental representation under either an $SU(2)\times U(1)$ or an $SO(4)$ group, turning off the SM gauge fields for now.  The main ingredients are 1) invariance under the $SU(2)\times U(1)$ or the $SO(4)$ group and 2) the Adler's zero condition in the scattering amplitudes of Goldstone bosons \cite{Adler:1964um}. It is possible to extend the approach we consider here to a general simple Lie group, which is discussed elsewhere \cite{Low:shift}. As a warm up exercise, we will first consider the trivial case of a single Goldstone boson resulting from spontaneously broken $U(1)$ symmetry.

\subsection{Spontaneously Broken $U(1)$}
\label{sect:u1coset}
In this case, it is well-known that the effective lagrangian can be constructed by imposing a constant shift symmetry,
\be
\label{eq:pishift}
\pi \to \pi + c \ ,
\ee
and the effective lagrangian is simply
\be
\label{eq:effLu1}
{\cal L} = \frac12 \partial_\mu \pi \partial^\mu \pi + {\cal O}(\partial^4)\ ,
\ee
where the shift symmetry manifests itself in the derivative coupling of the Goldstone boson. In other words, the Adler's zero condition is equivalent to imposing the shift symmetry in Eq.~(\ref{eq:pishift}).

Our approach, when it comes to more complicated symmetry breaking pattern, is to consistently require that the lagrangian be reduced down to Eq.~(\ref{eq:effLu1}) when turning off all but one flavor of Goldstone boson.

\subsection{$SU(2)\times U(1)$: No Custodial Symmetry}
\label{sect:nocus}

We assume  there is a doublet scalar $H=(h^1+i h^2, h^3+i h^4)/\sqrt{2}$ charged under the unbroken $SU(2) \times U(1)$. The nonlinear shift symmetry is postulated to be, 
\bea
\label{eq:Hshift}
H&\mapsto& H'=H+ \epsilon   +\frac1{f^2}\left[(\epsilon^\dagger H+ H^\dagger\epsilon)H -2 (H^\dagger H)\epsilon\right]\left[\ \sum_{n=0}^{\infty} \frac{a_n}{f^{2n}} \left(H^\dagger H\right)^n\right]  \nonumber \\
   &&\qquad  \qquad-\frac1{f^2}(\epsilon^\dagger H- H^\dagger\epsilon)H \left[\ \sum_{n=0}^{\infty} \frac{A_n}{f^{2n}} \left(H^\dagger H\right)^n\right]  \ ,
\eea
where $\epsilon=(\epsilon^1+i\epsilon ^2,\epsilon ^3+i \epsilon^4)/\sqrt{2}$ and we have kept only terms to linear order in $\epsilon$. The form of the higher order terms in the shift symmetry operation is dictated by the requirement that, when turning off all but one component, the shift symmetry reduces to the trivial shift symmetry for a single Goldstone boson in Eq.~(\ref{eq:pishift}). This is how we ensure the Adler's zero condition continues to be fulfilled even in the presence of the higher order terms in Eq.~(\ref{eq:Hshift}).

More specifically, at order $1/f^2$, there are three $SU(2)\times U(1)$ invariants one can build out of $\epsilon$ and $H$,\footnote{Notice that one cannot use the anti-symmetric tensor to construct the invariant $H^T \, i\sigma^2 \,\epsilon$, which is forbidden by the $U(1)$ symmetry carried by $H$ and $\epsilon$.}
\be
\label{eq:generaladler}
\frac1{f^2}\left[a (\epsilon^\dagger H)H + b (H^\dagger\epsilon)H + c  (H^\dagger H)\epsilon\right] \ .
\ee
The Adler's zero condition on $h^1$ amounts to the requirement that, when setting $h^2=h^3=h^4=\epsilon^2=\epsilon^3=\epsilon^4=0$, Eq.~(\ref{eq:generaladler}) reduces to Eq.~(\ref{eq:pishift}) for the $h^1$ component,
\be
h^1 \to h^1 + \epsilon^1 \ .
\ee 
This imposes the following condition
\be
a+b+c=0 \ .
\ee
Therefore there are two degrees of freedom that are not fixed by the Adler's condition, which we choose to be
\be
(a,b,c) = (1,1,-2) \quad {\rm and} \quad (1,-1,0) \ .
\ee
By $SU(2)$ invariance, keeping only $h^2$ by setting all other components to zero does not generate new conditions among the coefficients. This exercise can be performed order-by-order in $1/f$, leading to the shift symmetry proposed in Eq.~(\ref{eq:pishift}).

To construct the effective lagrangian, CCWZ instructs us to look for the Goldstone covariant derivative ${\cal D}_\mu H$ and the associated gauge field ${\cal E}_\mu$. Under the action of the nonlinear shift symmetry in Eq.~(\ref{eq:pishift}),  ${\cal D}_\mu H$ and ${\cal E}_\mu$ transforms homogeneously and non-homogeneously by a field-dependent $SU(2)\times U(1)$ rotation, respectively,
\bea
\label{eq:su2main}
{\cal D}_\mu H &\mapsto& {\cal D}_\mu H'= e^{i Y {\varphi}/f} \  e^{i \frac{u^a}f \frac{\sigma^a}2} \ {\cal D}_\mu H \ ,\\
 {\cal E}_\mu^a \frac{\sigma^a}2&\mapsto& e^{i \frac{u^a}f \frac{\sigma^a}2}\ {\cal E}_\mu^a \frac{\sigma^a}2\ e^{-i \frac{u^a}f \frac{\sigma^a}2} -i e^{-i \frac{u^a}f \frac{\sigma^a}2}\ \partial_\mu e^{i \frac{u^a}f \frac{\sigma^a}2}\ , \\
{\cal E}_\mu^Y&\mapsto& {\cal E}_\mu^Y - i e^{-i {\varphi}/f}\ \partial_\mu e^{i {\varphi}/f}\ ,
\eea
where $Y=1/2$ is the hypercharge of the Higgs doublet,  $\sigma^a, a=1,2,3$ are the Pauli matrices, and $\varphi$ and $u^a$ are the field-dependent $U(1)$ and $SU(2)$ phases, respectively.

Let's consider the Goldstone covariant derivative first, whose form is postulated to be 
\bea
\label{eq:DH}
{\cal D}_\mu H&=&  \partial_\mu H +\frac1{f^2}\left[(\partial_\mu H^\dagger H+H^\dagger\partial_\mu H)H -2 (H^\dagger H)\partial_\mu H\right]\left[\ \sum_{n=0}^{\infty} \frac{b_n}{f^{2n}} \left(H^\dagger H\right)^n\right] 
 \nonumber \\
   &&\quad -\frac1{f^2}(\partial_\mu H^\dagger H- H^\dagger\partial_\mu H)H \left[\ \sum_{n=0}^{\infty} \frac{B_n}{f^{2n}} \left(H^\dagger H\right)^n\right] \ .
\eea
Similar to Eq.~(\ref{eq:pishift}), this form is constrained by the linear $SU(2)\times U(1)$ invariance and the Adler's zero condition, which implies ${\cal D}_\mu H$ in Eq.~(\ref{eq:DH}) reduces to the Goldstone covariant derivative for a broken $U(1)$, ${\cal D}_\mu H \to \partial_\mu h^1$, upon setting all but $h^1$ to zero.  Since the Goldstone covariant derivative for a broken $U(1)$ is simply the ordinary derivative acting on the Goldstone field $\pi$,
\be
{\cal D}_\mu \pi \to \partial_\mu \pi \ ,
\ee
which is {\em invariant} under the shift symmetry in Eq.~(\ref{eq:pishift}), we see that the Adler's zero condition requires both $\varphi$ and $u^a$ to vanish when setting all components but $h^1$ to zero. These considerations lead to the following postulates, again working only  to linear order in $\epsilon$,
\bea
\label{eq:phi}
\varphi&=&\frac1{f} ( \epsilon^\dagger H- H^\dagger\epsilon)\left[\ \sum_{n=0}^{\infty} \frac{C_n}{f^{2n}} \left(H^\dagger H\right)^n\right] \ , \\
\label{eq:ua}
u^a&=&\frac1{f}( \epsilon^\dagger \frac{\sigma^a}2 H- H^\dagger \frac{\sigma^a}2 \epsilon) \left[\ \sum_{n=0}^{\infty} \frac{D_n}{f^{2n}} \left(H^\dagger H\right)^n\right]  \nonumber \\
      &&  +\frac1{f^3}(H^\dagger \frac{\sigma^a}2 H)( \epsilon^\dagger H- H^\dagger\epsilon)\left[\ \sum_{n=0}^{\infty} \frac{E_{n+2}}{f^{2n}} \left(H^\dagger H\right)^{n}\right] \ .
\eea
The remaining task now is to solve, order by order in $1/f$, for the Goldstone covariant derivative by plugging  Eqs.~(\ref{eq:Hshift}), (\ref{eq:DH}), (\ref{eq:phi}) and (\ref{eq:ua}) into Eq.~(\ref{eq:su2main}).

It turns out that this procedure is sufficient to obtain a unique solution for ${\cal D}_\mu H$, up to the overall normalization of the decay constant $f$. For example, all numerical coefficients up to order $1/f^{10}$ can be expressed in terms of a single parameter $a_0$,
\bea
(a_0, a_1, a_2, a_3, a_4)&=& (a_0,\ -\frac25 a_0^2 ,\ -\frac8{35} a_0^3,\  -\frac{24}{175} a_0^4,\  -\frac{32}{385} a_0^5 ) \nonumber \\
(A_0, A_1, A_2, A_3, A_4)&=& (3a_0,\ 6 a_0^2 ,\ \frac{72}{5} a_0^3,\  \frac{1224}{35} a_0^4,\  \frac{2976}{35} a_0^5 ) \nonumber \\
(b_0, b_1, b_2, b_3, b_4)&=& (-\frac12 a_0,\ \frac3{20} a_0^2 ,\ -\frac3{140} a_0^3,\  \frac{1}{560} a_0^4,\  -\frac{3}{30800} a_0^5 ) \nonumber \\
(B_0, B_1, B_2, B_3, B_4)&=& (\frac32 a_0,\ \frac9{4} a_0^2 ,\ \frac{27}{20} a_0^3,\  \frac{51}{112} a_0^4,\  \frac{279}{2800} a_0^5 ) \nonumber \\
(C_0, C_1, C_2, C_3, C_4)&=& \frac1{Y}(-\frac{9i}2 a_0,\ -9i a_0^2 ,\ -\frac{108i}{5} a_0^3,\  -\frac{1836i}{35} a_0^4,\  -\frac{4464i}{35} a_0^5 ) \nonumber \\ 
(D_0, D_1, D_2, D_3, D_4)&=& (-6i  a_0,\ -3i a_0^2 ,\ -\frac{9i}{5} a_0^3,\  -\frac{153i}{140} a_0^4,\  -\frac{93i}{140} a_0^5 ) \nonumber \\ 
(E_2, E_3, E_4, E_5)&=& ({9i} a_0^2,\  27i a_0^3 ,\ \frac{1377i}{20} a_0^4,\  \frac{4743i}{28} a_0^5 ) \nonumber \ .
\eea
In fact, it is not difficult to see that the series  resums into a compact expression for the Goldstone covariant derivative,
\bea
\label{eq:Dhsu2}
{\cal D}_\mu H &=&  \partial_\mu H+ \frac1{2 H^\dagger H}\left(1-\frac{\sin \frac{\sqrt{H^\dagger H}}{{f}}} {\frac{\sqrt{H^\dagger H}}{{f}}}\right)\left[ (H^\dagger \partial_\mu H+ \partial_\mu H^\dagger H) H -2(H^\dagger H) \partial_\mu H\right]\nonumber \\
&& - \frac{{f}}{(H^\dagger H)^{3/2}} \sin \frac{\sqrt{H^\dagger H}}{{f}} \ \sin^2 \frac{\sqrt{H^\dagger H}}{2{f}}\  (H^\dagger \partial_\mu H- \partial_\mu H^\dagger H) H \ ,
\eea
where we have absorbed the only free parameter $a_0$ into the decay constant ${f}\to f/\sqrt{6a_0}$. It is clear that $a_0$ represents the overall normalization of the decay constant $f$, which may vary from coset to coset. Otherwise, the Goldstone covariant derivative for $H$ is unique and universal among all possible symmetry breaking pattern $G/H$. Furthermore, from the bottom-up perspective nothing restricts ${f}$ to be a real number. Indeed, an imaginary ${f}$ gives rise to the effective lagrangian for a non-compact coset, while a real ${f}$ corresponds to a compact coset.

The above expression for the Goldstone covariant derivative depends on the particular basis chosen for the unbroken group generators. However, the resulting effective lagrangian is basis-independent. For example, the leading two-derivative lagrangian is given by,
\bea
\label{eq:l2}
{\cal L}^{(2)} &=& ({\cal D}_\mu H)^\dagger{\cal D}_\mu H \nonumber \\
  &=& \partial_\mu H^\dagger \partial^\mu H   \nonumber \\
  &&  + \frac{c_H}{2{f}^2} \partial_\mu(H^\dagger H) \partial^\mu(H^\dagger H)  +  \frac{c_r}{2{f}^2}  H^\dagger H \partial_\mu H^\dagger  \partial^\mu  H+ \frac{c_T}{2{f}^2}\left| \partial H^\dagger \,  H -H^\dagger \partial H\right|^2 \ , 
 \eea
 where
 \bea 
\label{eq:ch}
c_H&=& \frac{{f}^2}{2 H^\dagger H}\left(1-\frac{\sin^2 \frac{\sqrt{H^\dagger H}}{{f}}} {\frac{{H^\dagger H}}{{f}^2}}\right) \\
&=& \frac16 - \frac1{45} \frac{H^\dagger H}{{f}^2} +\frac1{630} \left(\frac{H^\dagger H}{ {f}^2}\right)^2+ {\cal O}\left(\frac{|H^\dagger H|^3}{{f}^6}\right)\ , \\
\label{eq:ct}
c_T&=& -2 \frac{{f}^4}{(H^\dagger H)^{2}} \sin^2 \frac{\sqrt{H^\dagger H}}{{f}} \ \sin^2 \frac{\sqrt{H^\dagger H}}{2{f}}\left(1-\sin^2 \frac{\sqrt{H^\dagger H}}{2{f}} \right) \\
&=&-\frac12 +\frac13 \frac{H^\dagger H}{{f}^2}-\frac1{10} \left(\frac{H^\dagger H}{ {f}^2}\right)^2+{\cal O}\left(\frac{(H^\dagger H)^3}{{f}^6}\right)\ , \\
\label{eq:cr}
c_r&=&-4c_H \ .
\eea
The lagrangian  ${\cal L}^{(2)}$ describes interactions of PNGBs in the fundamental representation of $SU(2)\times U(1)$ in, for examples,  both the $SU(3)/SU(2)\times U(1)$ coset \cite{Giudice:2007fh} and the littlest Higgs model based on the $SU(5)/SO(5)$ coset \cite{ArkaniHamed:2002qy}, up to the normalization of the decay constant $f$. Notice the $c_T$ term that violates the custodial invariance and contributes to the $\rho$ parameter in electroweak precision measurements.

It is worth commenting that ${\cal L}^{(2)}$ sits in a different operator basis from the one adopted by SILH, which manually sets $c_r=0$ by performing a field re-definition.

Now gauging the $SU(2)_L\times U(1)_Y$ is a simple procedure of replacing the ordinary derivative by the gauge covariant derivative,
\be
\partial_\mu \to D_\mu = \partial_\mu - i\, g \frac{\sigma^a}{2} W^a_\mu -i\, g^\prime Y B_\mu \ ,
\ee
in the $c_r$ and $c_T$ term in Eq.~(\ref{eq:l2}). Since the derivative in the $c_H$ term acts on a gauge singlet, $H^\dagger H$, it remains an ordinary derivative. Turning on a background field for the CP-even neutral component of the doublet,  $ H \to (0, h)^T/\sqrt{2}$, the two-derivative effective lagrangian becomes
\be
\label{eq:hequation}
\frac12 \partial_\mu h\partial^\mu h + \frac14 g^2 (2f^2) \sin^2 \frac{h}{\sqrt{2}f}\left( W^{+\,\mu} W_\mu^- + \frac{1}{2\cos^2\theta_w} Z_\mu Z^\mu\right) \ ,
\ee
If we define in the SM $v\approx 246$ GeV such that $m_W^2 = g^2 v^2/4$, we see
\be
\label{eq:hvev}
v^2 = 2f^2  \sin^2 \frac{\langle h\rangle }{\sqrt{2}f} \ ,
\ee
where $\langle h\rangle$ is the vacuum expectation value (VEV) of $h$.

Next we consider the object that transforms in-homogeneously under $SU(2)_L\times U(1)_Y$ like a gauge field:
\bea
{\cal E}_\mu^a \frac{\sigma^a}2&\mapsto& e^{i \frac{u^a}f \frac{\sigma^a}2}({\cal E}_\mu^a \frac{\sigma^a}2) e^{-i \frac{u^a}f \frac{\sigma^a}2} -i e^{-i \frac{u^a}f \frac{\sigma^a}2}\partial_\mu e^{i \frac{u^a}f \frac{\sigma^a}2} \\
{\cal E}_\mu^Y&\mapsto& {\cal E}_\mu^Y - i e^{-i {\varphi}/f} \partial_\mu e^{i {\varphi}/f}
\eea
Their forms are postulated to be
\bea
{\cal E}_\mu^a&=&\frac1f( \partial_\mu H^\dagger\frac{\sigma^a}2 H- H^\dagger\frac{\sigma^a}2 \partial_\mu H) \left[\ \sum_{n=0}^{\infty} \frac{G_n}{f^{2n}} \left(H^\dagger H\right)^n\right]  \nonumber \\
      &&  +\frac1f(H^\dagger\frac{\sigma^a}2 H)( \partial_\mu H^\dagger H- H^\dagger\partial_\mu H)\left[\ \sum_{n=0}^{\infty} \frac{I_{n+1}}{f^{2n}} \left(H^\dagger H\right)^{n}\right] \ , \\
 {\cal E}_\mu^Y &=&     \frac1f (\partial_\mu H^\dagger H- H^\dagger\partial_\mu H)\left[\ \sum_{n=0}^{\infty} \frac{J_n}{f^{2n}} \left(H^\dagger H\right)^n\right] \ , 
 \eea
Again, after solving for the series order by order in $1/f$, compact forms can be obtained,  
\bea
{\cal E}_\mu^a&=& i\frac{\sqrt{2}}{H^\dagger H}\left(-1+\cos \frac{\sqrt{H^\dagger H}}{f}\right)( \partial_\mu H^\dagger\frac{\sigma^a}2 H- H^\dagger \frac{\sigma^a}2 \partial_\mu H) \nonumber \\
 &&\qquad -i \frac{2\sqrt{2}}{(H^\dagger H)^2}\sin^4 \frac{\sqrt{H^\dagger H}}{2f} (H^\dagger \frac{\sigma^a}2 H)( \partial_\mu H^\dagger H- H^\dagger\partial_\mu H) \\
 {\cal E}_\mu^Y&=& \frac{3i}{8 Y}\frac1{H^\dagger H}\sin^2\frac{\sqrt{H^\dagger H}}{f}(\partial_\mu H^\dagger H- H^\dagger\partial_\mu H) \ ,
 \eea
where we have again re-scaled  $f\to f \sqrt{6a_0}$. In the above $Y$ is the hypercharge of the Higgs doublet. If we choose $Y=\sqrt{3/2}$, $ {\cal E}_\mu^Y$ gives the corresponding result for $SU(3)/SU(2)\times U(1)$, which is the charge of the $SU(2)$ doublet under the $U(1)$ subgroup of $t^8={\rm Diag}(1,1,-2)/\sqrt{6}$.

 \subsection{$SO(4)$: With Custodial Symmetry}
 
Since $SO(4)$ is a simple Lie group, one can apply the general formulas in Ref.~\cite{Low:shift}. However, it is instructive to adopt the same "ad hoc" approach as in the semi-simple group of $SU(2)\times U(1)$. In this case the Higgs transforms as a fundamental representation, $\vec{h}=(h_1, h_2, h_3, h_4)^T$, of $SO(4)$. We will use the basis of $SO(4)$ generators in Ref.~\cite{Low:2009di} such that $H=(h_1+i h_2, h_3+i h_4)^T/\sqrt{2}$. Similar to the $SU(2)\times U(1)$ case, we write
 \bea
 {\cal D}_\mu\vec{h} &=&  \partial_\mu \vec{h} - (\vec{h}\cdot \vec{h}\ \partial_\mu\vec{h} - \vec{h}\cdot\partial_\mu\vec{h}\ \vec{h})  \left[\ \sum_{n=0}^{\infty} \frac{B_n}{f^{2n}} \left(\vec{h}\cdot \vec{h}  \right)^n\right]   \ , \\
 u^A&=& \vec{\epsilon}^{\ T} T^A \vec{h}  \left[\ \sum_{n=0}^{\infty} \frac{C_n}{f^{2n}} \left(\vec{h}\cdot \vec{h}  \right)^n\right]  \ , \\
 {\cal E}^A &=&\partial_\mu \vec{h}^{\ T} T^A \vec{h}  \left[\ \sum_{n=0}^{\infty} \frac{D_n}{f^{2n}} \left(\vec{h}\cdot \vec{h}  \right)^n\right]  \ ,
 \eea
 where $T^A$ are the $SO(4)$ generators. The action of the shift symmetries is
 \bea
 \vec{h} &\mapsto& \vec{h}+\vec{\epsilon}- (\vec{h}\cdot \vec{h}\  \vec{\epsilon} - \vec{h}\cdot\vec{\epsilon}\ \vec{h}) \left[\ \sum_{n=0}^{\infty} \frac{A_n}{f^{2n}} \left(\vec{h}\cdot \vec{h}  \right)^n\right]   \ , \\
 {\cal D}_\mu\vec{h} &\mapsto&  e^{i \frac{u^A}f T^A} {\cal D}_\mu\vec{h} \ , \\
 {\cal E}_\mu^A T^A&\mapsto& e^{i \frac{u^a}f T^A}({\cal E}_\mu^a T^A) e^{-i \frac{u^a}fT^A} -i e^{-i \frac{u^a}f T^A}\partial_\mu e^{i \frac{u^a}f T^A} \ .
 \eea
 In the end we have
 \bea
  {\cal D}_\mu\vec{h} &=& \frac1f \partial_{\mu} \vec{h} +   \frac1{f \vec{h}\cdot\vec{h}}\left(1- \frac{f\sqrt{2}}{\sqrt{\vec{h}\cdot\vec{h}}}  \sin \frac{\sqrt{\vec{h}\cdot\vec{h}}}f \right) ( \vec{h}\cdot\partial_\mu\vec{h}\ \vec{h}- \vec{h}\cdot\vec{h} \ \partial_\mu\vec{h}) \ , \\
   {\cal E}_\mu^A&=& \frac{2i}{\vec{h}\cdot\vec{h}}\left(-1+\cos \frac{\sqrt{\vec{h}\cdot\vec{h}}}{\sqrt{2}f} \right)\partial_\mu \vec{h}^{T} T^A \vec{h} \ ,
 \eea  
from which the entire effective lagrangian can be built.
 
It is interesting to compare the $SO(4)$ results with the case without the $SU(2)_C$ invariance derived in Sect.~\ref{sect:nocus}. Re-writing in $SU(2)\times U(1)$ notation we find
\be
 {\cal D}_\mu\vec{h}=  \partial_\mu H+ \frac1{2 H^\dagger H}\left(1-\frac{\sin \frac{\sqrt{H^\dagger H}}{f}} {\frac{\sqrt{H^\dagger H}}{f}}\right)\left[ (H^\dagger \partial_\mu H+ \partial_\mu H^\dagger H) H -2(H^\dagger H) \partial_\mu H\right]\ , 
 \ee
 which is identical to Eq.~(\ref{eq:Dhsu2}) upon setting to zero the coefficient of the custodial-violating operator, $c_T=0$.

\section{Yukawa couplings}

Any realistic model must contain Yukawa couplings giving rise to fermion masses in the SM. In turn, the Yukawa couplings necessarily break the shift symmetries acting on the Higgs boson, since they generate one-loop contributions to the Higgs mass.\footnote{The gauge sector and the scalar potential in the SM also break the shift symmetry and generate one-loop quadratic divergences to the Higgs mass. However, these are  subleading contributions to the top quark loop. Therefore in many models the gauge and scalar contributions remain un-cancelled. They can be addressed in a similar fashion as the Yukawas and will not be considered further in this work.}  As is well-known, the quadratically divergent contribution from the top quark would  destabilize a light Higgs boson if it is not cancelled at one-loop. The cancellation can be achieved if we require that  the shift symmetry is  broken softly, only by relevant operators.

In what follows we give a simple implementation of the top Yukawa sector. The starting point is the usual third-generation electroweak doublet $q_L=(u_L, d_L)^T$ and singlet $d_R$. A vector-like pair of electroweak singlet fermions, $(U_L, U_R)$, with charge $Q=2/3$, are also introduced. The Yukawa lagrangian is, to all orders in $1/f$,
\be
{\cal L}_y= \left(\sum_{n=0}^{\infty} \lambda_{2n} \frac{|H^\dagger H|^n}{f^{2n}}\right)  f\, \bar{u}_RU_L + \left(\sum_{n=0}^{\infty} \lambda_{2n+1} \frac{|H^\dagger H|^n}{f^{2n}}\right)\,  \bar{u}_R H^\dagger  q_L+ \lambda_Uf \, \bar{U}_R U_L + {\rm h. c.}  \ ,
\ee
where the $\lambda_U$ term parameterizes the soft-breaking of the shift symmetry. Focusing on the top-like fermions, $(u_L, u_R)$ and $(U_L, U_R)$, the top Yukawa coupling is given by  
\be
\label{eq:topyukawa}
{\cal L}_t= \left(\sum_{n=0}^{\infty} \lambda_{2n} \frac{h^{2n}}{ 2^n f^{2n}}\right) f \, \bar{u}_R U_L + \left(\sum_{n=0}^{\infty} \frac{\lambda_{2n+1}}{\sqrt{2}} \frac{h^{2n+1}}{2^nf^{2n}}\right)  \bar{u}_R u_L + \lambda_U f\, \bar{U}_R U_L + {\rm h. c.} \ .
\ee
We will postulate that, under the shift symmetry acting on the uneaten neutral component $h\to h+\epsilon$, the fermions transform at linear order in $\epsilon$ by
\bea
u_L &\mapsto& u_L - i \alpha \, \frac{\epsilon}{f} U_L \ , \qquad
U_L \mapsto U_L - i \alpha \frac{\epsilon}{f} u_L \ , \label{eq:fermiont}\\
u_R &\mapsto& u_R \ , \qquad\qquad \qquad\ 
U_R \mapsto U_R \ , \nonumber
\eea
where $\alpha$ is the "charge" of the fermions $u_L$ and $U_L$ under the shift symmetry. In essence, $(u_L, U_L)$ transform into each other under the shift symmetry acting on $h$. In order to suppress the top contribution to the Higgs mass,  the shift symmetry transformation in Eq.~(\ref{eq:fermiont}) can only be broken softly, by the $\lambda_U$ term in ${\cal L}_t$, which lead to the following recursion relation for $\lambda_n$,
\be
i \sqrt{2} \alpha \,\lambda_n = (n+1) \lambda_{n+1} \ .
\ee
It is then easy to see that the series in Eq.~(\ref{eq:topyukawa}) resums into
\be
{\cal L}_t= \lambda \, f\, \bar{u}_RU_L  \cos\frac{\alpha h}f + i \lambda \,f\,   \bar{u}_R   u_L \sin\frac{\alpha h}f+ \lambda_U f\, \bar{U}_R U_L + {\rm h. c.}  \ ,
\ee
where $\lambda\equiv \lambda_0$.

The top-like fermion mass matrix in the basis $(u_R, U_R)$ and $(u_L, U_L)$ is
\be
{\cal M}=\left( \begin{array}{cc}
 i\lambda f \sin(\alpha h/f) &\lambda f \cos(\alpha h/f) \\
0 & \lambda_U f
        \end{array} \right)\ ,
\ee
from which we solve for the mass eigenvalues
\bea
m_t &=& \frac{\lambda_t}{\sqrt{2}} h \left[ 1 -\frac{\alpha^2 (\lambda_U^4-\lambda_U^2\lambda^2+\lambda^4)}{6(\lambda_U^2+\lambda^2)^2}\frac{h^2}{f^2} +{\cal O}\left(\frac{h^4}{f^4}\right) \right] \ , \\
m_T &=& f\sqrt{\lambda_U^2+\lambda^2} \left[1+ \frac{\lambda_t^2}{4(\lambda_U^2+\lambda^2)} \frac{h^2}{f^2} +{\cal O}\left(\frac{h^4}{f^4}\right)\right] \ ,
\eea
where
\be
\lambda_t = \frac{\alpha \lambda_U\lambda}{\sqrt{\lambda_U^2+\lambda^2}} \ .
\ee
In the end the SM top quark has mass $m_t$ after the uneaten neutral component gets a VEV, $\langle h\rangle$, which is related to the SM VEV, $v\approx 246$ GeV, through Eq.~(\ref{eq:hvev}).

We can check that a Higgs mass is not generated at the one-loop level through the Coleman-Weinberg potential \cite{Coleman:1973jx} by computing the trace of ${\cal M}^\dagger{\cal M}$,
\be
{\rm Tr}\, {\cal M}^\dagger{\cal M} = f^2(\lambda^2 +\lambda_U^2) \ .
\ee
On the other hand, the determinant of ${\cal M}^\dagger{\cal M}$ enters into the Higgs couplings to massless SM gauge bosons such as gluons and photons through the Higgs low-energy theorem \cite{Ellis:1975ap,Shifman:1979eb} and is calculated to be
\be
 {\rm Det}\, {\cal M}^\dagger{\cal M} = \lambda^2 \lambda_U^2 f^4 \, \sin^2 \frac{\alpha h}{f} \ ,
 \ee
which is independent of the mass eigenvalues and has a trigonometric form that was observed previously in Refs.~\cite{Low:2010mr,Falkowski:2007hz}. It is easy to convince oneself that this will always be the case when the shift symmetry is broken only softly {\em and} when the fermions carry the same nonlinear charge $\alpha$ under the shift symmetry. If some of the fermions carry different charges, say $\alpha_1$ and $\alpha_2$, then the determinant will be a function of $\sin \alpha_1 h/f$ and $\sin\alpha_2 h/f$.  Similar observations in coset constructions can be found in Refs.~\cite{Azatov:2011qy,Montull:2013mla}.

Generally speaking, in models with a PNGB Higgs, the top Yukawa coupling is responsible for generating the scalar potential for the Higgs boson. The Higgs mass obtained in this way typically is very close to the top mass and, therefore, too heavy for the observed 125 GeV. This can be remedied by additional model-building efforts or allowing some degree of tuning in the scalar potential.
     
\section{Phenomenological constraints and collider signals}

In this section we briefly consider phenomenological constraints and collider signals of a minimally symmetric Higgs.\footnote{A detailed study is outside of the scope of the present work and will be pursued elsewhere.} In this regard it is convenient to assume the custodial symmetry and consider the low-energy effective lagrangian introduced in Ref.~\cite{Contino:2010mh},
\be
\label{eq:leff}
{\cal L}_{eff} = \frac12 \partial_\mu h \partial^\mu h - V(h) +\frac{v^2}{4} {\rm Tr}\left( D_\mu\Sigma^\dagger D^\mu \Sigma\right)\left[ 1+ 2a \frac{h}v + b\frac{h^2}{v^2} + b_3 \frac{h^3}{v^2}+\cdots \right] \ .
\ee
The $2\times 2$ matrix $\Sigma$ is defined as
\be
\Sigma = e^{i\sigma^a \pi^a(x)/v} \ ,
\ee
where $\pi^a(x)$ are the eaten Goldstone bosons  that became the longitudinal components of the $W$ and $Z$ bosons. The coefficients $a, b$, and $b_3$ can be computed by matching Eq.~(\ref{eq:leff}) to the lagrangian for the minimally symmetric Higgs in  Eq.~(\ref{eq:hequation}), where the coefficient of the $W^+W^-$ term gives
\bea
&&\frac14 g^2(2f^2) \sin^2\frac{h+\langle h\rangle}{\sqrt{2}f}  \nonumber \\
&=& \frac14 g^2(2f^2) \left[\sin^2\frac{\langle h\rangle }{\sqrt{2}f} + \frac{h}{\sqrt{2}f}\sin \frac{2\langle h\rangle }{\sqrt{2}f} + \frac{h^2}{2f^2} \cos \frac{2\langle h\rangle }{\sqrt{2}f} - \frac{1}{3\sqrt{2}} \frac{h^3}{f^3} \sin \frac{2\langle h\rangle }{\sqrt{2}f} + \cdots \right] \ .
\eea
After expressing $\langle h\rangle$ in terms of the SM VEV, $v\approx 246$ GeV, using Eq.~(\ref{eq:hvev}), we arrive at
\be
\label{eq:abb3}
a=\sqrt{1-\frac{v^2}{2f^2}} \ , \qquad b = 1-\frac{v^2}{f^2} \ , \qquad b_3= -\frac23 \frac{v^2}{f^2}\sqrt{1-\frac{v^2}{2f^2}} \ .
\ee
The SM limit is obtained by taking $f$ to infinity, giving rise to $a^{\rm (SM)}=b^{\rm (SM)}=1$ and $b_3^{\rm (SM)}=0$. The parameter $a$ enters into the Higgs couplings to two electroweak gauge bosons, and is constrained by both direct measurements of Higgs properties, which is rather weak, as well as from precision electroweak measurements, which constraint on $a$ at the 95\% confidence level is \cite{Contino:2013gna,Ciuchini:2013pca},
\be
\label{eq:aconstraint}
0.98 \le a \le 1.12 \ ,
\ee
when one assumes no other new physics contributions to precision electroweak measurements. It should be stressed that, in absence of other means to determine the normalization of the decay constant $f$, one cannot use the above equation to constrain $f$. Instead, one should use $a$ to fix the normalization of $f$ and proceed to make other predictions to compare with experiments. For example, one can use Eq.~(\ref{eq:aconstraint}) to derive a constraint on $b$,
\be
0.92 \le b \le 1.50 \ ,
\ee
at 95\% C.L., which is an unambiguous constraint on a minimally symmetric Higgs, independently of the coset structure. Experimentally $b$ can be measured by performing the scattering process, $W_LW_L\to hh$, at an energy above the mass scale of other resonances appearing in the $W_LW_L$ scattering. For a study at a future high energy linear collider see Ref.~\cite{Contino:2013gna}.

Similarly, when considering smoking gun signals of a minimally symmetric Higgs we need to focus on observables that are independent of the normalization of $f$. In this regard, we again turn to vector boson scattering where $a$ and $b$ enter as follows \cite{Contino:2010mh},
\bea
{\cal A}(W_LW_L\to Z_LZ_L) &\approx& \frac{s}{v^2}(1-a^2)\ ,  \\  
{\cal A}(W_LW_L\to hh)&\approx&  \frac{s}{v^2}(b-a^2)\ .
\eea
In the above $s$ is the partonic centre-of-mass energy and subleading contribution in $m_W^2/s$ have been dropped. In the limit that the 125 GeV Higgs completely unitarizes the vector boson scattering, $a=b=1$ and there is no growth with respect to energy in the amplitudes. In a minimally symmetric Higgs, the complete unitarization never occurs.
Plugging Eq.~(\ref{eq:abb3}) in the amplitudes above we see that 
\be
1-a^2=-(b-a^2)=v^2/(2f^2)\ ,
\ee
leading to  
\be
\label{eq:predict}
{\cal A}(W_LW_L\to Z_LZ_L)  = -{\cal A}(W_LW_L\to hh) = \frac{s}{2f^2} + {\cal O}(s^0) \ .
\ee
Being a prediction of a minimally symmetric Higgs, the above equation is universal among all PNGB Higgs models, regardless of the coset employed. It would be interesting to consider the reach of this measurement in a future high energy $pp$ collider.\footnote{In addition to $Z_LZ_L$ and $hh$ final states, it is conceivable that similar relations involving $Z_Lh$ final state could exist. This is left for future work.}

Another universal prediction in vector boson scattering is the triple Higgs production through $V_LV_L\to hhh$, $V=W, Z$, where \cite{Contino:2013gna}
\be
{\cal A}( V_LV_L\to hhh) = \frac{is}{v^3}(4ab -4a^3-3b_3) \ .
\ee
Again plugging in Eq.~(\ref{eq:abb3}) the above amplitude vanishes identically,
\be
{\cal A}( V_LV_L\to hhh)=0 \ .
\ee
That is the triple Higgs production vanishes in the high energy limit, which has been previously proved in a symmetric coset $G/H$ in Ref.~\cite{Contino:2013gna} using a $Z_2$ symmetry of the Lie algebra. Here we see the prediction is  slightly more general, since what is needed is the $Z_2$ symmetry acting on  the $SU(2)\times U(1)$ sector, which nonetheless can be embedded in a non-symmetric coset.

It should be stressed that the predictions considered in this section arise directly from the nonlinear sigma model lagrangian that respects the shift symmetry. There could exist new gauge bosons and scalars that would show up as resonances in scatterings of longitudinal gauge bosons, which are highly model-dependent. Therefore, the universal predictions can only be checked by performing the scattering at an energy scale above the resonances, which would require a future high energy collider.

\section{Conclusion}
Without recourse to the CCWZ formalism, we assume four shift symmetries acting on the four real components of the Higgs doublet, which, together with linearly realized electroweak $SU(2)_L\times U(1)_Y$ symmetry, is sufficient to derive the effective lagrangian, up to the overall normalization of the Goldstone decay constant $f$. As such, the effective lagrangian is universal among models with a PNGB Higgs, regardless of the symmetry breaking pattern $G/H$, up to terms that explicitly violate the shift symmetries. The universality of the effective lagrangian, up to the overall normalization of decay constant $f$, can be checked explicitly using the survey done in Refs.~\cite{Low:2010mr,Alonso:2014wta}.

There exist models with more than one Higgs doublet, where additional scalars such as a second doublet or an $SU(2)$ triplet scalars might exist. In these cases
our effective lagrangian describes self-interactions of the PNGB within the $SU(2)$ doublet. It is also possible to derive an effective lagrangian for the $SU(2)$ triplet scalars that preserve the shift symmetries, which is outside of the scope of present work. (However, see Ref.~\cite{Low:shift}.)

The conventional coset construction of PNGB Higgs models can be viewed as UV completions of a minimally symmetric Higgs boson. The fact that we used only the Adler's zero condition and the linearly realized $SU(2)_L\times U(1)_Y$ symmetry to derive the effective lagrangian suggests that self-interactions of Goldstone bosons are universal in the IR, being independent of the details of symmetry breaking in the UV. The one parameter that cannot be determined in the IR is the normalization of the Goldstone decay constant $f$, which is sensitive to the UV coset.

The Yukawa interactions, on the other hand, necessarily break the shift symmetries. However, by demanding the symmetry is broken only softly by relevant operators, it is possible to reduce the UV sensitivity of the Higgs mass arising from the top sector. Adding only a vector-like pair of electroweak singlet fermions, we gave a simple implementation of the top Yukawa interactions where there is no one-loop quadratic divergence in the Higgs mass from the top sector.

Using the effective lagrangian we derived, it is possible to make predictions that are universal among all PNGB models. In particular, we showed that vey high energy vector boson scatterings, above the scale of new resonances, could potentially carry smoking gun signals of a minimally symmetric Higgs. In this regard it will be interesting to study the possibility of making such measurements in a future high energy collider.

\begin{acknowledgments}

The author is grateful for  insightful conversations with Nima Arkani-Hamed.  This work was supported in part by the U.S. Department of Energy under Contracts No. DE-AC02-06CH11357 and No. DE-SC0010143, and was initiated at KITP in Santa Barbara, which is supported by the U.S. National Science Foundation under Grant No. NSF PHY11-25915. Hospitality at the Center for Future High Energy Physics at IHEP in Beijing is acknowledged, where part of this work was completed.

\end{acknowledgments}


\end{document}